\documentclass[letterpaper,aps, prl, reprint, superscriptaddress,footinbib]{revtex4-1}
\usepackage[T1]{fontenc}
\PassOptionsToPackage{utf8}{inputenc}
\usepackage{color}
\usepackage{xcolor}
\usepackage{pdfcolmk}
\usepackage{verbatim}
\usepackage{mathtools}
\usepackage{amsmath}
\usepackage{amssymb}
\usepackage{eucal}
\usepackage[normalem]{ulem}
\usepackage{bbm, dsfont}
\usepackage{float}
\usepackage{array}
\usepackage{mathrsfs}
\usepackage{braket}
\usepackage{textcomp}

\makeatletter

\providecolor{lyxadded}{rgb}{0,0,1}
\providecolor{lyxdeleted}{rgb}{1,0,0}

\DeclareRobustCommand{\lyxsout}[1]{\ifx\\#1\else\sout{#1}\fi}

\usepackage[colorlinks,citecolor=blue,linkcolor=red]{hyperref}

\begin{document}
\title{Giant and robust Josephson diode effect in multiband topological nanowires}

 \author{Bao-Zong Wang}
 \thanks{The authors make equal contributions.}
 \affiliation{International Center for Quantum Materials, School of Physics, Peking University, Beijing 100871, China}
 \affiliation{Hefei National Laboratory, Hefei 230088, China}

\author{Zi-Kai Li}
 \thanks{The authors make equal contributions.}
 \affiliation{International Center for Quantum Materials, School of Physics, Peking University, Beijing 100871, China}

 \author{Zhong-Da Li}
\affiliation{International Center for Quantum Materials, School of Physics, Peking University, Beijing 100871, China}
 \affiliation{Hefei National Laboratory, Hefei 230088, China}

\author{Xiong-Jun Liu}
\thanks{Corresponding author: xiongjunliu@pku.edu.cn}
\affiliation{International Center for Quantum Materials, School of Physics, Peking University, Beijing 100871, China}
\affiliation{Hefei National Laboratory, Hefei 230088, China}
\affiliation{International Quantum Academy, Shenzhen 518048, China}

\begin{abstract}
We theoretically predict the giant and robust Josephson diode effect in quasi-one-dimensional topological Majorana nanowires in the regime with multiple subbands, which is expected to be relevant for the real experiment. In the multiband regime, the Majorana bound states and conventional Andreev bound states naturally coexist, and respectively contribute to the fractional and conventional parts in the Josephson effect, with the former/latter having 4$\pi$/2$\pi$-periodicity. We show that the interplay between the two types of bound modes can produce a robust and giant diode effect in the deep topological phase regime. Notably, we unveil a novel spin parity exchange mechanism, occurring only in the multiband regime, which leads to a robust high efficiency plateau of the giant diode effect. This effect is a nontrivial consequence of the balanced Fermi moment shifts of the multiple subbands in tuning the external magnetic field. Our finding highlights the multiband engineering as a powerful tool to optimize the Josephson diode effect realistically and provides a feasible signature to identify topological phase regime in real superconducting nanowires.
\end{abstract}

\maketitle

\textcolor{blue}{\em Introduction.}--
The Majorana bound states (MBSs) have attracted significant interest for obeying non-Abelian statistics, and can serve as a cornerstone for fault tolerant quantum computing~\cite{alicea2012new,KITAEV20032,2008_RevModPhys,2001_PRL_Ivanov,ying2020non,PhysRevLett.119.047001}. A leading platform for realizing MBSs is the proximitized nanowire~\cite{nanowire_PRL_2010_DasSarma,Nanowire_PRL_2010_Oreg,Nanowire_PRB_2010_DasSarma,2021_nanowire,2021_PRB_Das_Sarma,2021_PRL_Das_Sarma,2024_CPSun_Majonara}, where a semiconductor wire with strong spin-orbit interaction (SOI) is coupled to a superconductor (SC) under magnetic field. This setup can host MBSs at two ends of the nanowire, with experimental signatures such as zero-bias conductance peaks~\cite{peak_2020_NRP,peak_2011_PRL,peak_2012_NL,peak_2012_NP,peak_2012_science,peak_2013_PRB} and the fractional $4\pi$-periodic Josephson effect~\cite{2004_LowTP_Kwon,2009_LiangFu_PRB,2004_EPJB,laroche2019observation,2016_nc_4piexper}. The latter provides a key tool for probing MBSs, reflecting their unique phase behavior in Josephson junctions. Despite significant progress~\cite{MBSexp_2018_Science,MBSexp_2016_Nature_exponential,MBSexp_2016_Science_Majorana,MBSexp_2025_Nature_microsoft,MBSexp_2019_Nature_JJ}, obtaining unambiguous confirmation of MBSs remains a persistent challenge that continues to drive ongoing research.

Diodes, which allow current to flow mainly in one direction, are essential in modern electronics. While conventional diodes suffer from Joule heating due to resistance, superconducting diodes provide a dissipation free alternative. This advantage has spurred rapid interest in the superconducting diode effect (SDE). The SDE is generally divided into intrinsic~\cite{SDE_2022_NC_Bauriedl,SDE_2020_Nature_Observation,SDE_2021_APE_Observation,SDE_2021_PNAS_LiangFu,SDE_2022_NJP_Nagaosa,SDE_2022_NN_Field-free,SDE_2022_PNAS_LiangFu,SDE_2022_PRL_Youichi,SDE_2022_PRR,SDE_2023_PRL,SDE_2024_Nature_Tanaka,SDE_2024_PRL_QFSun,SDE_2024_pan_superconducting,SDE_2024_wan_unconventional} and Josephson diode effects (JDE)~\cite{JDE_2007_PRL_DaiXi,JDE_2021_PRB_Nagaosa,JDE_2022_arxiv_CongjunWu,JDE_2022_LiangFu,JDE_2022_Nature,JDE_2022_NN_Denis,JDE_2022_NP,PhysRevB.112.014509,JDE_2022_PRX_JiangKun}, with the latter drawing particular attention. When both inversion and time-reversal symmetries are broken~\cite{2013_liu_PRB_Manipulating,2012_Liu_ABSsIn1DTSC}, the forward ($I_c^+$) and reverse ($I_c^-$) directional critical currents are generally unequal~\cite{2013_liu_PRB_Manipulating}, a key mechanism underlying the SDE. 
In JDE, the Andreev bound states (ABSs) in the junction region contribute to the supercurrent~\cite{ABS_2018,ABS_2018_Mizushima,ABS_2018_PRL_Measurement,ABS_2019_PRX,ABS_PRL_2020}. Recent experiments have observed the JDE in various systems~\cite{ciaccia2023gate,banerjee2023phase,matsuo2023josephson,trahms2023diode,pal2022josephson,baumgartner2022supercurrent,diez2023symmetry,wu2022field}. Alongside these experimental advancements, theoretical frameworks have been developed for the JDE behavior~\cite{PhysRevB.107.184511,doi:10.1126/sciadv.abo0309,hu2007proposed,PhysRevB.103.245302,PhysRevLett.131.096001,PhysRevLett.129.267702,PhysRevB.105.104508,PhysRevB.106.214504,PhysRevB.106.134514,PhysRevB.103.144520,gupta2023gate,PhysRevB.108.214519,PhysRevApplied.18.034064,PhysRevB.106.214524,PhysRevB.107.245415,PhysRevLett.130.266003,PhysRevB.103.245302,PhysRevLett.130.177002,PhysRevLett.129.267702,PhysRevResearch.5.033199,PhysRevB.108.174516,Costa2024JDE,fukaya2024JDEvortex,Debnath_2025_FieldFree,JDE_2025_debnath,yerin2025supercurrent}, of particular interest is the JDE realized in topological nanowires hosting MBSs. Previous studies have focused on ideal one-dimensional (1D) models~\cite{2024_PRB_JingWang_Ideal1DJDE,DanielLoss_Parity,mondal2025josephson}, whereas realistic systems with multiple subbands~\cite{2021_PRB_Das_Sarma, 2021_PRL_Das_Sarma, liu2012zero} have been largely overlooked. Such systems may exhibit distinct mechanisms that can significantly modify the diode efficiency.

\begin{figure}[t]
\centering
\includegraphics[width=1.0\columnwidth]{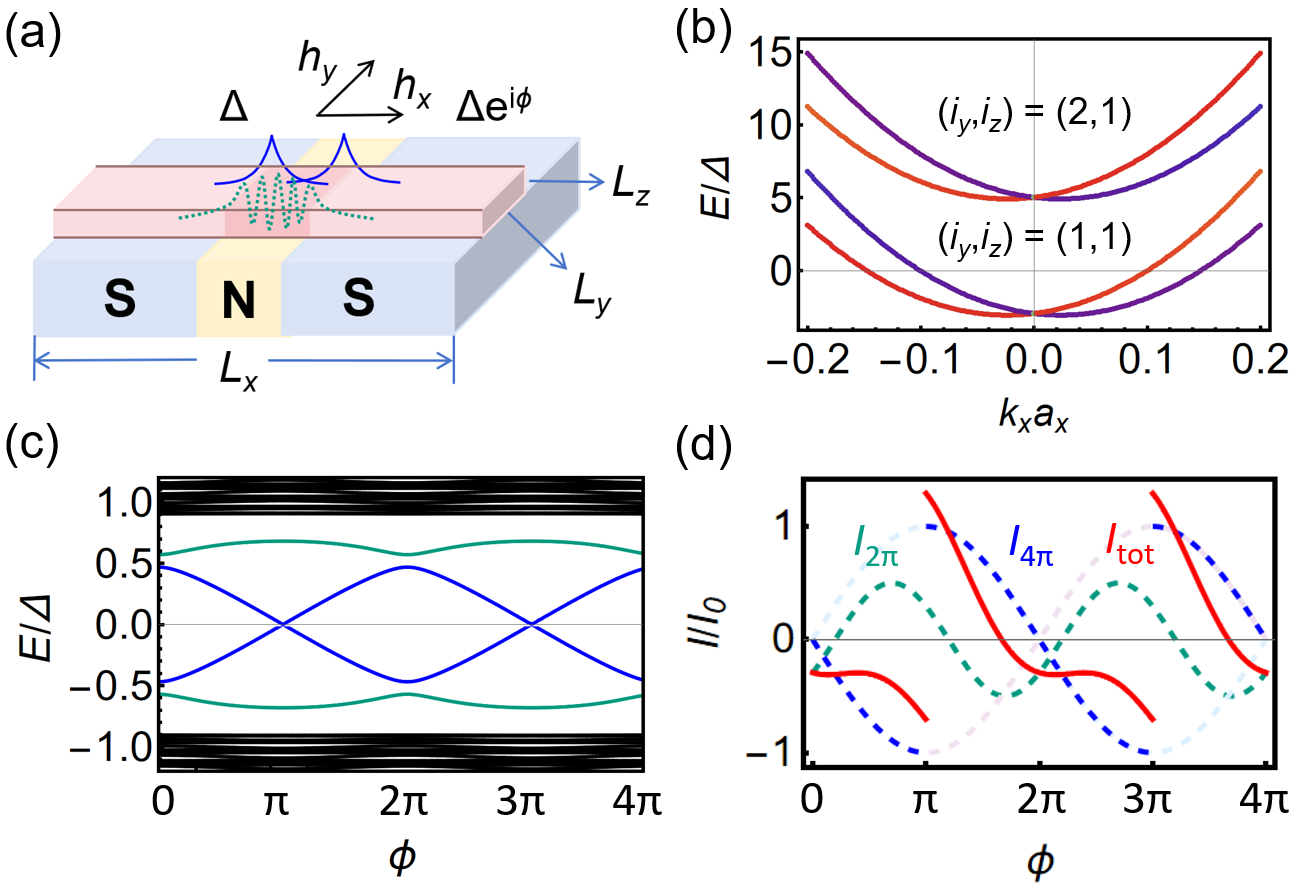}
\caption{\label{FIG.1} (a) Quasi-1D Josephson junction nanowire on an $s$-wave SC, with MBSs (blue) and ABSs (light-green). (b) The lowest energy spinful subbands with $(i_y,i_z)$ = (1,1) and (2,1) are shown. (c) Energy spectrum in the deep topological region showing isolated MZMs (blue), coupled MZMs (green), and bulk states (black). (d) Interplay of $2\pi$- and $4\pi$-periodic Josephson currents. Dashed blue ($I_{4\pi}$) and green ($I_{2\pi}$) curves denote ground-state MBS and ABS contributions, respectively. The discontinuity in the total current $I_{\mathrm{tot}}$ arises from an abrupt parity switch at the crossing point of the $4\pi$-periodic current $I_{4\pi}$.
}
\end{figure}

In this Letter, we predict a giant and robust JDE, together with a novel mechanism uncovered here, for Majorana nanowires in the multiband regime. The multiband topological nanowire hosts both MBSs and ABSs, which we show produce a large JDE by balancing the contributions from the fractional and conventional Josephson effects, 
with the diode efficiency remaining substantial in the deep topological phase.
Importantly, we uncover a novel mechanism that the spin-parity band exchange among subbands leads to a robust plateau of high diode efficiency 
as tuning the external magnetic field. 
Our findings emphasize the critical role of the spin-parity subband engineering in optimizing the JDE, and provide a realistic method to identify topological phases for MBSs.

\textcolor{blue}{\em The multiband model.}--We start with the multiband model for a single quasi-1D nanowire placed on the $s$-wave SC utilizing the proximity effect, 
forming a Josephson junction as depicted in Fig.~\ref{FIG.1}(a). 
The system has finite dimensions $L_x$, $L_y$, $L_z$ along the three spatial directions with trapping potentials. In typical nanowire material, the system is strongly confined along the $y$ and $z$ directions, and we consider the rectangular cross section with a width of $L_y\sim10^2\,\text{nm}$ and a thickness of $L_z\sim10^0\, \text{nm}$, resulting in a reasonable approximation of $L_z \ll L_y \ll L_x$. Consequently, we limit our analysis to the lowest eigenstates in the $z$-direction, setting $N_z = 1$~\cite{2021_PRB_Das_Sarma,2021_PRL_Das_Sarma}. As for the $y$-direction, we focus on cases that a few transverse eigenmodes in this direction can be occupied (e.g. with the number of orbits $N_y \le 3$, see Fig.~\ref{FIG.1}(b)).
The resulting effective lattice Hamiltonian takes the form
$H_{\mathrm{nw}} = H_0+H_{\text{SOI}}+H_B+H_\Delta$, whose explicit form is given by
\begin{align}
& H_0 = -t_x \sum_{\boldsymbol{i}} (c_{\boldsymbol{i}+\boldsymbol{\delta}_x}^{\dagger} c_{\boldsymbol{i}} + \text{h.c.})-\sum_{\boldsymbol{i}} (\mu_{\boldsymbol{i}}-2t_x) c_{\boldsymbol{i}}^{\dagger} c_{\boldsymbol{i}}, \nonumber\\
& H_{\text{SOI}} = \mathrm{i} \sum_
{\boldsymbol{i}} \left[ c_{\boldsymbol{i}+\boldsymbol{\delta}_x}^{\dagger} (\alpha_x\hat{\sigma}_y) c_{\boldsymbol{i}} -c_{\boldsymbol{i}+\boldsymbol{\delta}_y}^{\dagger} (\alpha_y\hat{\sigma}_x)c_{\boldsymbol{i}} \right]  + \text{h.c.}, \nonumber\\
& H_B = \sum_{\boldsymbol{i}} c_{\boldsymbol{i}}^{\dagger} (h_x\hat{\sigma}_x + h_y\hat{\sigma}_y) c_{\boldsymbol{i}},  \nonumber\\
& H_\Delta = \sum_{\boldsymbol{i}} \left( \Delta_{\boldsymbol{i}} c_{\boldsymbol{i} \uparrow}^{\dagger} c_{\boldsymbol{i}\downarrow}^{\dagger} + \text{h.c.} \right),
\end{align}
with electron creation operator denoted by the spinor \(c_{\boldsymbol{i}}^\dagger = (c_{\boldsymbol{i}\uparrow}^\dagger, c_{\boldsymbol{i}\downarrow}^\dagger)\) and $\boldsymbol{i}=(i_x,i_y)$. Note that the index $i_x$ denotes the real site position along the $x$-direction, while $i_y=1,2,...,N_y$ is an effective position. 
The Hamiltonian consists of four contributions: The free-particle term $H_{0}$ describes nearest-neighbor hopping $t_x$ along $\boldsymbol{\delta}_x=(1,0)$ and includes an $i_y$-site-dependent (actually transverse-mode dependent) chemical potential 
$\mu_{\boldsymbol{i}}=\mu_0 - i_y^2(\pi \hbar)^2/(2m^* L_y^2)$~\cite{Suppl}, where the energy of transverse modes is approximated by that in a 1D infinite square well~\cite{2021_PRL_Das_Sarma}, 
and $m^*$ is the effective mass of the electron.
The SOI term $H_{\text{SOI}}$ incorporates the longtitutional SOI with intensity $\alpha_x$ and the effective transverse SOI between different orbits with intensity $\alpha_y \propto L_y^{-1}$, which is nonzero only when $\boldsymbol{\delta}_y$=(0,1)~\cite{Suppl}, with the Pauli matrices $\hat{\sigma}_i (i=x,y,z)$ on spin space. 
The term $H_B$ describes a uniform magnetic field $\boldsymbol{h}=(h_x,h_y)$ that applied throughout the entire nanowire, where $h_x$ opens a Zeeman gap in the spectrum, while $h_y$ shifts the Fermi surface along the $x$ direction, breaking inversion symmetry~\cite{2013_liu_PRB_Manipulating,pnas.2119548119}.
Finally, $s$-wave pairing is depicted by $H_{\Delta}$, with $\Delta_{\boldsymbol{i}}$ nonzero only in the superconducting leads, carrying a phase difference $\phi$ across the junction: $\Delta_{\mathrm{R}}=\Delta_{\mathrm{L}} e^{i\phi}=\Delta e^{i\phi}$. 

\begin{figure}[t]
\includegraphics[width=1.0\columnwidth]{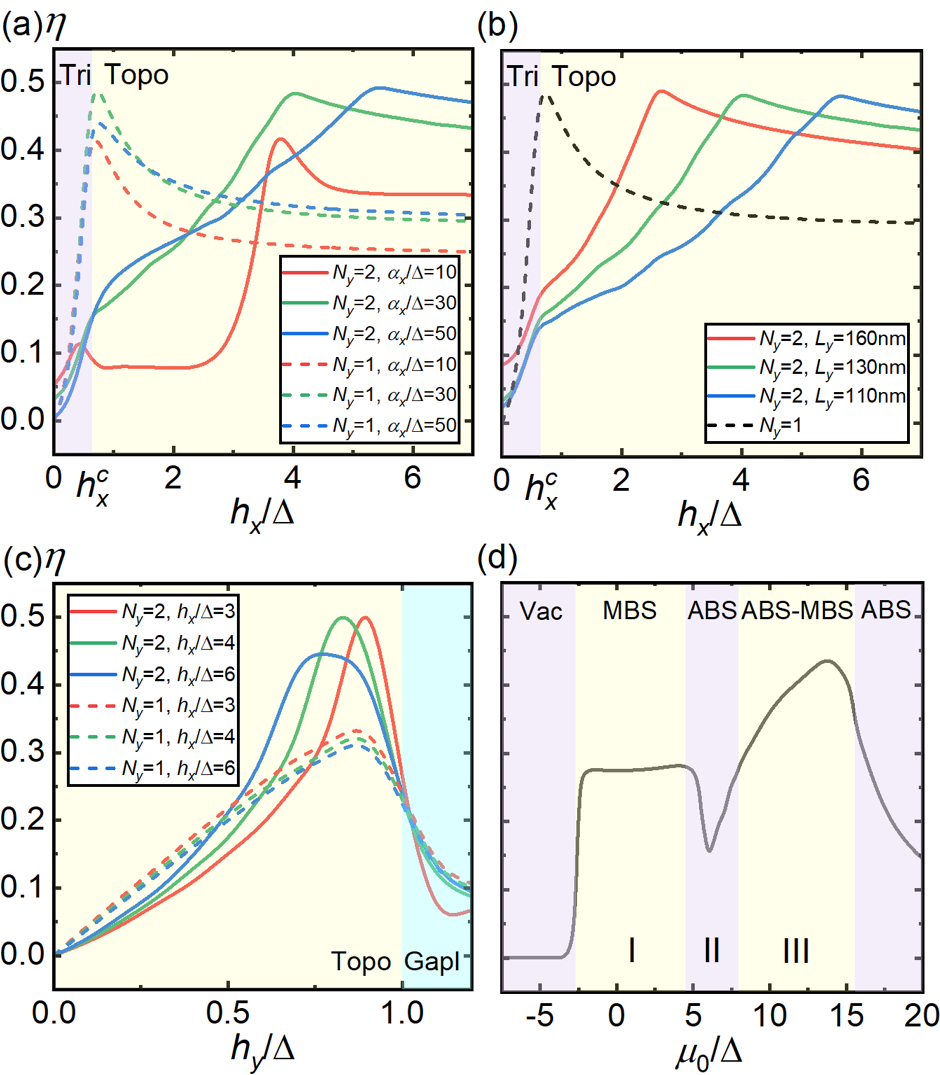}
\caption{\label{FIG.2} Relatively high and stable efficiency $\eta$ of the system populating 3 spinful subbands in the region of strong $h_x$ under various parameters. (a) $\eta$ vs $h_x$ for typical $\alpha_x$'s when 1 subband (calculated with $N_y=1$, ideal 1D model) or 3 subbands (evaluated with $N_y=2$) are occupied. 
(b) $\eta$ vs $h_x$ for realistic $L_y$'s. (c) $\eta$ vs $h_y$ for different $h_x$'s. (d) $\eta$ vs $\mu_{0}$ when other parameters are fixed. ``Tri'', ``Topo'', ``Gapl'' and ``Vac'' stand for ``Trivial'', ``Topological'', ``Gapless'' and ``Vacuum'', respectively. Parameters: $\alpha_x$ is set to be 30 in panel (b)-(c), and $h_y=0.8$ in panel (a)-(c). In panel (d), $\alpha_x=10,\,h_x=5,\, h_y=0.7$.}
\end{figure}

\textcolor{blue}{\em Balance between fractional and conventional Josephson currents.}--
In the multiband regime, the Fermi level intersects several subbands. When the number of Fermi points is odd, the system is topological with MBSs, while the conventional ABSs can coexist at the nanowire ends [Fig.~\ref{FIG.1}(a)]. In a Josephson junction, MBSs couple and contribute a fractional Josephson effect with a $4\pi$ periodicity~\cite{2004_LowTP_Kwon,2009_LiangFu_PRB,2004_EPJB,laroche2019observation,2016_nc_4piexper}, whereas the conventional ABSs lead to $2\pi$-periodic Josephson currents~\cite{ABS_2018,ABS_2018_Mizushima,ABS_2018_PRL_Measurement,ABS_2019_PRX,ABS_PRL_2020}. These two components, denoted $I_{4\pi}$ and $I_{2\pi}$ [Fig.~\ref{FIG.1}(d)], respectively, coexist and together determine the total supercurrent at zero temperature as a function of the junction phase difference $\phi$. The total current reads~\cite{PhysRevLett.67.3836}

\begin{equation}
    I(\phi) = \frac{2e}{\hbar}\frac{\partial E}{\partial\phi},\, E=-\frac{1}{2}\sum_{E_{i} \ge0}E_{i}(\phi).
\end{equation}
where the sum runs over all positive-energy eigenstates.
Fig.~\ref{FIG.1}(d) displays the total current, $I_{\text{tot}}=I_{4\pi}+I_{2\pi}$ (red line), which develops a global maximum distinct from its minimum. Consequently, the critical currents become asymmetric, $I_c^+ \neq I_c^-$, with $I_c^{\pm}=\max[\pm I_{\text{tot}}(\phi)]$. This asymmetry constitutes direct evidence of the Josephson diode effect, characterized by the efficiency $\eta=(I_c^+ - I_c^-)/(I_c^+ + I_c^-)$. In this multiband regime, the JDE originates from the competition between the fractional current $I_{4\pi}$ and the conventional component $I_{2\pi}$. 

We adopt experimentally realistic parameters for InAs and InSb nanowires proximitized by Nb or Al superconductors~\cite{2021_PRL_Das_Sarma}: with $L_y = 130\,\text{nm}$, $m^* = 0.04m_e$,  $\alpha_x = 0.1 \,\text{eV}\cdot\mathring{\text{A}}$ and $\Delta = 2.4\,\text{K}$ ($0.2\,\text{meV}$). Taking the effective lattice spacing along the $x$ direction to be $a_x = 4\,\mathrm{nm}$, the corresponding tight-binding parameters are $t_x = 200\Delta$, $\alpha_x=6\Delta$, $\alpha_y=1.3\Delta$, and let $\Delta=1$ for convenience. The chemical potential $\mu_0$ is set such that the Fermi level crosses an odd number of subbands, ensuring the emergence of MBSs. Moreover, the Fermi level is set to lie midway between the two eigenmodes at $k_x = 0$.
We consider the short junction with a normal region length of $N_\text{N}=2$.
The Josephson current is evaluated using the recursive Green's function method~\cite{Sunqingfeng2016PRB,Sunqf2024PRL}, which efficiently avoids the computational cost in exact diagonalization for large system sizes 
\begin{equation}
I=-e\left\langle\frac{\mathrm{d}\hat{N}_{\text{L}}}{\mathrm{d}t}\right\rangle =\frac{e}{\hbar} \mathrm{Tr} \left\{ \Gamma_z \left[ {G}_{\text{10}}^<(t,t) \hat{T}-\hat{T}^\dagger {G}_{\text{01}}^<(t,t) \right] \right\}.
\end{equation}
Here, the particle number on the left lead is defined as $\hat{N}_{\text{L}} = \sum_{i_x\leq0,i_y}\sum_{\sigma=\uparrow,\downarrow}c_{\boldsymbol{i}\sigma}^{\dagger} c_{\boldsymbol{i}\sigma}$, with $\Gamma_z=\sigma_z\otimes I_{N_y\times N_y}$ and $\hat{T}$ is the hopping matrix connecting the left superconductor to the normal region, $G_{\text{NL}}^<(t,t)$ is the lesser Green's function.
For simplicity, we often set $e=\hbar=1$.

Fig.~\ref{FIG.2} shows the diode efficiency versus $h_x$ for different system parameters like the SOI strength $\alpha_x$ (Fig.~\ref{FIG.2}(a)), the nanowire's transverse length $L_y$ (Fig.~\ref{FIG.2}(b)), and $h_y$ (Fig.~\ref{FIG.2}(c)). For the single band regime (see the colored dashed lines), we can find that the efficiency $\eta$ consistently exhibits a pronounced peak near the critical topological phase transition boundary $h_x^c=\sqrt{\Delta^2-h_y^2}$~\cite{2024_PRB_JingWang_Ideal1DJDE} and diminishes if tuning into deep topological and trivial regimes. In the trivial (topological) regime, the JDE are primarily contributed from the higher harmonic terms of conventional (fractional) Josephson currents $I_{2\pi}$ ($I_{4\pi}$) due to the ABSs (MBSs)~\cite{DanielLoss_Parity}, for which the diode effect is weak. 
Only near the phase transition boundary $h_x\approx h_x^c$, which enables a competition between the two types of Josephson currents, an enhanced JDE is resulted.

\begin{figure}[t]
\includegraphics[width=1.0\columnwidth]{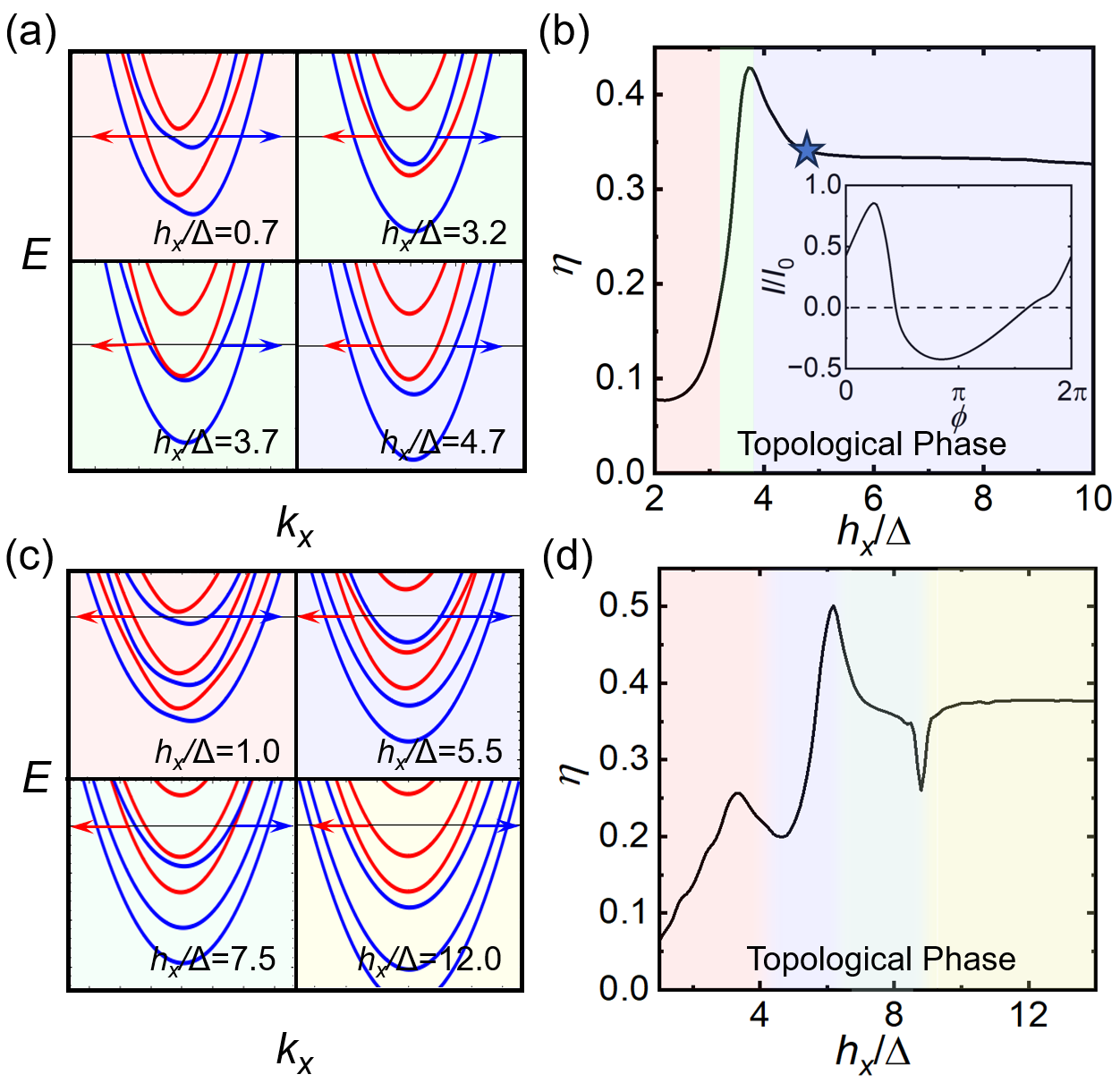}
\caption{\label{Fig3} High diode efficiency plateau by spin-parity subband exchanges mechanism. (a) Energy bands before, during, and after the spin-parity band exchange in the 2-orbit ($N_y=2$) model with 3 occupied subbands. Colored bands denote different spin-parity subbands, and colored arrows mark the Fermi point shifts induced by the inversion-symmetry-breaking field $h_y$. (b) Diode efficiency $\eta$ as a function of $h_x$, where the background colors correspond to the band structures shown in (a). The current-phase relation at the marked point ($h_x = 4.8\Delta$) is shown in the insets. (c) Energy bands showing the spin-parity band exchange in the three-orbital ($N_y=3$) model with five occupied subbands. (d) $\eta$ versus $h_x$, with color segments matching the band structures in (c). Parameters: $\alpha_x=10,\, h_y=0.8$.}
\end{figure}

The multiband regime shows a novel new feature that the strong JDE is generically obtained in the deep topological phase regime. 
Within the multiband topological phase, the contributions from ABSs and MBSs coexist, leading to a persistent competition between $I_{4\pi}$ and $I_{2\pi}$ across the entire topological regime. Consequently, although the maximum position of diode efficiency varies depending on system parameters, the regime of high diode efficiency extends over most of the topological phase region [Fig.~\ref{FIG.2}(a)-(c)].

In Fig.~\ref{FIG.2}(d), increasing the chemical potential drives the Fermi level across one, two, and three subbands, defining regimes I, II, and III. Region $\mathrm{I}$ features a low efficiency platform contributed from MBSs and region $\mathrm{II}$ exhibits weak JDE dominated by ABSs. In contrast, in region $\mathrm{III}$ a high diode efficiency is driven by the coexistence of MBSs and ABSs.



\textcolor{blue}{\em Spin-parity band exchange mechanism.}--A more intriguing discovery is that we predict a robust high efficiency plateau with a novel spin-parity band exchange mechanism obtained only in the multiband regime.
The spin-parity $\mathbb{P}_s$ of a certain subband is defined as the sign of the eigenvalue of the spin-coupled term of the Hamiltonian $h_x\hat{\sigma}_x + (h_y+2\alpha_x\sin k_xa_x)\hat{\sigma}_y$ at $k_y=0$, where the interband SOI vanishes. As illustrated in Fig.~\ref{Fig3}, the bands with $\mathbb{P}_s = +1$ and $\mathbb{P}_s = -1$ are plotted in red and blue color, respectively. The relative positions of normal subbands (with $\Delta=0$) with different spin parities can be tuned by $h_x$ and shows a crucial impact on the diode effect. As $h_x$ increases, the subbands with opposite spin parities shift oppositely in energy, leading to the spin parity exchange of the two types of subbands when $h_x$ exceeds a threshold. We consider first the topological phase for $N_y=2$, with three normal subbands crossing Fermi energy [Fig.~\ref{Fig3}(a)]. We can see that at $h_x = 0.7\Delta$ with $h_y=0.8\Delta$, on Fermi level there are two blue bands with $\mathbb{P}_s=-1$ and one red band with $\mathbb{P}_s=+1$. At $h_x\approx|\mu_1-\mu_2|/2 \approx 3.5\Delta$, the upper two subbands begin to swap positions, and the full exchange is completed at $h_x\geq4.7\Delta$. Then the spin parity configuration is robust for further increasing $h_x$.
The diode efficiency in Fig.~\ref{Fig3}(b) highlights three distinct regions corresponding to the band exchange shown on the left. Prior to the band exchange, the diode efficiency remains to be low. At the band exchange region, a sharp enhancement in the diode efficiency emerges. After the complete exchange of the two subbands with different spin parities, a robust and high JDE plateau is obtained. 

The predicted large and robust JDE is not limited to $N_y=2$, but a generic result applicable to the multiband regime. 
To demonstrate this, we further consider $N_y = 3$, where the topological phase is realized by tuning the Fermi level to intersect five subbands, comprising two red and three blue bands [Fig.~\ref{Fig3}(c)]. With increasing $h_x$, two spin parity exchange transitions are obtained, respectively at $h_x\approx6.2\Delta$ and $h_x\approx8.8\Delta$. Following the second transition, once the spin-parity configuration becomes stabilized, a robust efficiency plateau generically develops [Fig.~\ref{Fig3}(d)]. 

\begin{figure}[t]
\includegraphics[width=1.0\columnwidth]{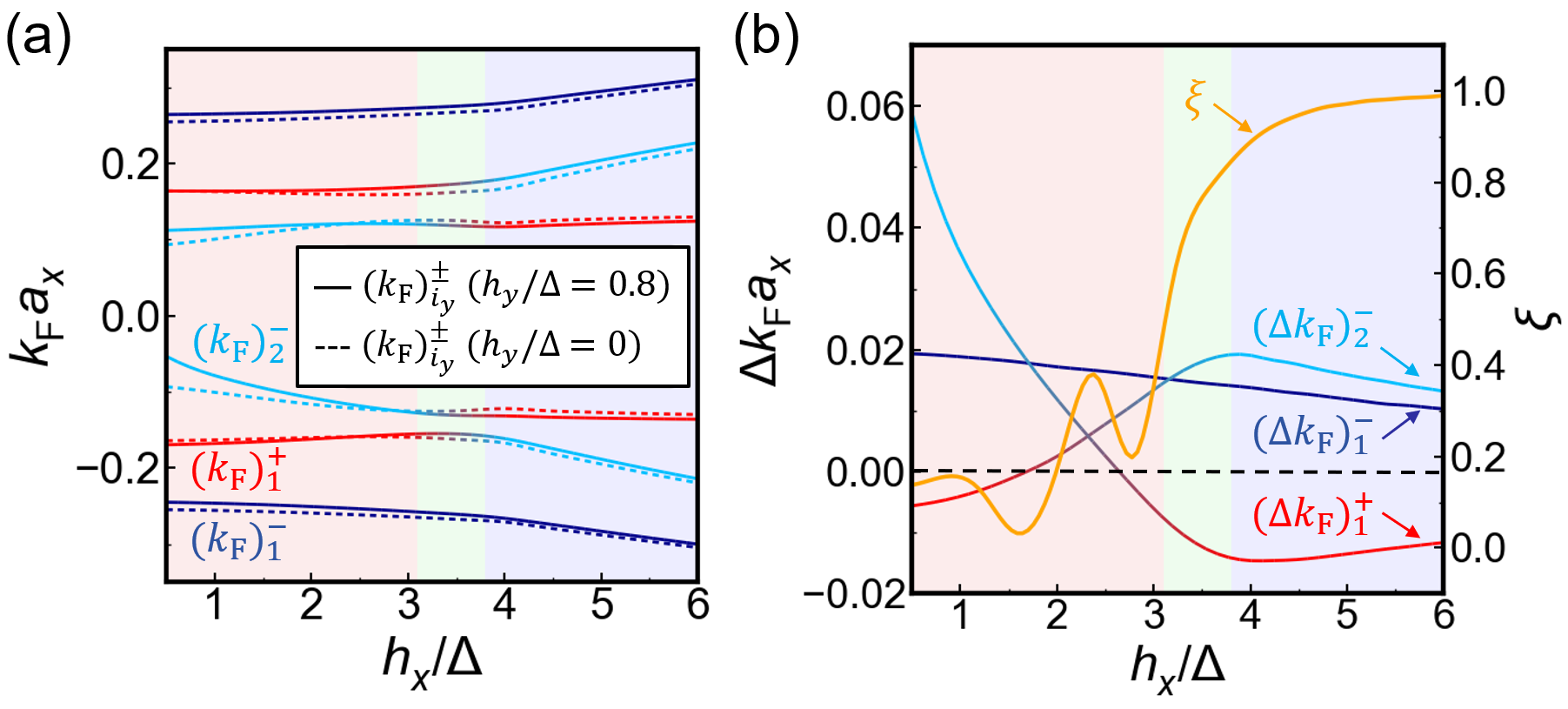}
\caption{\label{Fig4} Underlying mechanism of high diode efficiency plateau after the spin-parity subband exchange. (a) Dependence of Fermi momenta $(k_\textrm{F})_{i_y}^{\pm}a_x$ on field strength $h_x$ for $N_y=2$ with three occupied subbands and the same parameters as Fig.~\ref{Fig3}(a).
(b) Fermi momentum shifts $\Delta k_\textrm{F}=k_\textrm{F}(h_y)-k_\textrm{F}(h_y=0)$. The yellow line denotes a dimensionless ratio $\xi$ that quantifies the competition between left- and right-shifted Fermi points. Parameters: $\alpha_x=10$, $h_y=0.8$. } 
\end{figure}
The giant efficiency plateau implies a novel mechanism featured by the spin parity configuration of the subbands. As illustrated in Fig.~\ref{Fig3}(a), the Fermi level intersects three subbands. For \( h_y = 0 \), the bands remain symmetric about \( k_x = 0 \). With $h_y\neq0$, the 1D dispersions take the form \( E_{i_y}^\pm(k_x) = 2t_x(1-\cos k_xa_x) - \mu_{i_y} \pm \sqrt{h_x^2+(h_y + 2\alpha_x \sin k_xa_x)^2}, \) 
with the orbit and spin-parity being labeled by $i_y$ and $\pm$, respectively. 
Therefore, upon applying $h_y$, the net shift of the two Fermi points in a given subband is
\begin{equation}
(\Delta k_\textrm{F})_{i_y}^\pm \approx \mp \frac{2}{\hbar |v_{\textrm{F}}|}{\left[1+ \left(\frac{h_x}{2\alpha_x \sin k_\textrm{F}a_x}\right)^2 \right]}^{-\frac{1}{2}}h_y.
\end{equation}
This expression quantifies the directional deformation of Fermi points, where the red bands ($\mathbb{P}_s=+1$) shift leftward and the blue bands ($\mathbb{P}_s=-1$) shift rightward [Fig.~\ref{Fig3}(a)].
Fig.~\ref{Fig4}(a,b) shows the Fermi point positions and their shifts for the three occupied bands with $N_y=2$ as a function of the magnetic field $h_x$. 
We find that $(\Delta k_{\textrm{F}})_1^+$ and $(\Delta k_{\textrm{F}})_2^-$ initially decrease, cross zero, and eventually saturate.
The zero crossing signifies an exchange of the Fermi points.
These Fermi point shifts are induced by the transverse field $h_y$. 
A key feature is that the dimensionless ratio $\xi=2|(\Delta k_\textrm{F})_1^+|/[(\Delta k_{\textrm{F}})_1^- + (\Delta k_{\textrm{F}})_2^-]$, which quantifies the relative shift between left- and right-shifted total Fermi momenta, saturates to a stable plateau after the spin-parity exchange [the yellow line in Fig.~\ref{Fig4}(b)].
This saturation signals the balance between the two types of Josephson currents and corresponds to the emergence of strong JDE. In this regime, the two blue bands with spin-parity $\mathbb{P}_s = -1$ exhibit nearly identical and constant Fermi point displacements, which mainly contribute
the conventional Josephson current $I_{2\pi}$. In contrast, the remaining red subband predominantly contributes to the fractional Josephson current $I_{4\pi}$ originated from the MBSs. 

We also note that this balance stems from the competition between the Josephson current originating from the higher-curvature red band (with lower chemical potential from band bottom) and that originating from the lower-curvature blue subbands (with higher chemical potential from band bottom). 
In general, through the spin-parity band exchange, we can combine the $N_y$ lower-curvature bands of identical spin parity to balance the $N_y-1$ opposite spin-parity bands with higher curvature.
Such a mechanism provides a novel generic scheme to optimize the JDE with high and robust efficiency in the multiband topological regime.

\textcolor{blue}{\em Conclusion.}--We have shown that multiband Majorana nanowires provide a realistic and robust platform for realizing the giant JDE. The coexistence of MBS and ABS in the multiband regime leads to a persistent competition between the fractional and conventional Josephson currents, sustaining large diode efficiency deep into the topological phase. Especially, we uncovered a novel spin-parity band exchange mechanism unique to the multiband regime, which realizes a robust high efficiency plateau. These results demonstrate subband engineering as a practical route to optimize SDE and offer a new probe for identifying topological phase in Majorana nanowire systems.

\textcolor{blue}{\em Acknowledgments.}-- The authors are thankful to Yue Mao and Yu-Hao Wan for fruitful discussions. This work was supported by the National Natural Science Foundation of China (Grants No. 12425401 and No. 12261160368), the National Key Research and Development Program of China (2021YFA1400900), Quantum Science and Technology-National Science and Technology Major Project (Grant No. 2021ZD0302000), and Shanghai Municipal Science and Technology Major Project (No.~2019SHZDZX01).

\bibliography{ref}

\renewcommand{\thesection}{S-\arabic{section}}
\setcounter{section}{0}  
\renewcommand{\theequation}{S\arabic{equation}}
\setcounter{equation}{0}  
\renewcommand{\thefigure}{S\arabic{figure}}
\setcounter{figure}{0}  
\renewcommand{\thetable}{S\Roman{table}}
\setcounter{table}{0}  
\onecolumngrid \flushbottom 

\newpage
\begin{center}
\large \textbf{\large Supplementary Material: Giant and robust Josephson diode effect in multiband topological nanowires}
\end{center}

\section{From quasi-1D nanowire to multiband Hamiltonian}

In this section, we derive the effective tight-banding multiband Hamiltonian from a realistic quasi-1D nanowire, and find proper values for parameters for numerical calculation. Specifically, we consider a semiconductor quantum well based on an InAs-Al/Sb heterostructure. As reported in Refs.~\cite{2021_PRB_Das_Sarma,2021_PRL_Das_Sarma}, it exhibits an effective electron mass of $m^* = 0.04\,m_e$, and a spin-orbit interaction (SOI) strength of approximately $(\alpha_x)_{\text{C}} = 0.1\,\text{eV}\cdot\mathring{\text{A}}$ (for the continuous model). Typical system dimensions are \( L_z < 10\,\text{nm} \), \( L_y \approx 130\,\text{nm} \), and \( L_x \) that extend over several microns, validating the approximation \( L_z \ll L_y \ll L_x \).

We first derive the hopping term $t_x$ and the spin-orbit interaction term $\alpha_x$ along the $x$-direction, which correspond to the parameters of an ideal one-dimensional nanowire. The horizontal lattice constant is set to be $a_x=4\,\text{nm}$. For a simple lattice system with only spin-independent hopping and Rashba SOI, it is crucial to ensure that both the continuous model (described by $m^*$ and $(\alpha_x)_{\text{C}}$) and the tight-banding model (with the nearest-neighbor hopping element $t_x$ and $(\alpha_x)_{\text{TB}}$) accurately capture the low-energy physics near the band bottom. To establish a clear correspondence between these descriptions, we expand the electronic dispersion around $k_x = 0$. By retaining terms up to the first-order in $k_x$, we extract the SOI parameter for the tight-binding model, while the hopping term is determined from the second-order expansion:
\begin{align}   (\alpha_x)_{\text{C}}k_x \approx2(\alpha_x)_{\text{TB}}\sin (k_xa_x),&\, \quad \alpha_x=(\alpha_x)_{\text{TB}}=\frac{(\alpha_x)_{\text{C}}}{2a_x}=1.2\,\text{meV}, \nonumber \\
\frac{\hbar^2k_x^2}{2m^*}\approx 2t_x-2t_x\cos(k_xa_x),&\, \quad t_x=\frac{\hbar^2}{2m^*a_x^2}=40\,\text{meV}.
\end{align}

Next, we estimate the parameters $\mu_{\boldsymbol{i}}=\mu_{i_y}$ and $\alpha_y$ to qualitatively describe multiple orbits. The variation in the chemical potential across different orbits arises from the confinement along the $y$-direction. For simplicity, we model the system in the $y$-direction as an infinite potential well, leading to the following energy spectrum and corresponding eigenstates:
\begin{equation}
    E(i_y)=i_y^2\cdot\frac{\pi^2\hbar^2}{2m^*L_y^2},\, \quad \psi_{i_y}(y) =\sqrt{\frac{2}{L_y}}\sin\frac{i_y\pi y}{L_y}.
\end{equation}
Here, \( i_y \) is a positive integer that labels the energy levels. These energy levels determine the variation in the chemical potential among the subbands, while the background chemical potential \( \mu \) can be uniformly tuned via a metallic gate in experiments:
\begin{equation}
    -\mu_{i_y}=-\mu_0+E(i_y)=-\mu_0+i_y^2\cdot(\frac{\pi a_x}{L_y})^2t_x.
\end{equation}

From the above, we can define \( V = E(1) = \pi^2 (a_x / L_y)^2 t_x = 0.5\,\text{meV}\) to reflect the characteristic energy difference between orbits in the $y$ dimension in the effective lattice model, with higher energy levels following \( E(i_y) = i_y^2\cdot E(n_y = 1)=i_y^2V \). 
Meanwhile, the SOI in the \( y \)-direction can be incorporated by the second quantization:
\begin{equation} 2\alpha_xa_x\braket{\psi_{i_y}|\partial_y|\psi_{i_y'}}=2\alpha_xa_x \int_0^{L_y} {\frac{2}{L_y}}\sin\frac{i_y\pi y}{L_y} \partial_y \sin\frac{i_y'\pi y}{L_y} \mathrm{d}y = \frac{2\alpha_xa_x}{L_y}A_{i_yi_y'}=\alpha_y \frac{A_{i_yi_y'}}{A_{21}}.
\end{equation}
One can calculate $\alpha_y=0.2\,\text{meV}$, and $A_{i_yi_y}=0$, $A_{21}=8/3$, $A_{31}=0$, $A_{32}=24/5$. 

To simplify numerical calculations, physical quantities with energy dimensions are rescaled to dimensionless units. Given the significant contribution of Majorana bound states to the supercurrent, it is convenient to set the BCS pairing strength $\Delta = 2.4\,\text{K} = 0.2\,\text{meV}$ as unity in computations. Consequently, the parameters are chosen as $t_x = 200$, $\alpha_x = 6$, $V = E(n_y=1) = 2.5$, and $\alpha_y = 1.3$, ensuring consistency with the precision of \( m^* \) and \( (\alpha_x)_{\text{C}} \). 
Considering possible candidate materials for the nanowire, one can adopt the SOI intensity of the same magnitude.


For the possible values of \( h_x \) and \( h_y \), we note that the effective Lande factor in an InSb-based semiconducting nanowire is \( g_{\text{eff}} \approx 50 \), with \( t_x \) and \( \alpha_x \) comparable to those in InAs. By adopting a dimensionless \( h \) and normalizing \( \Delta \) to unity in the computation, the corresponding magnetic field in SI units is given by \( B = \Delta / (g_{\text{eff}} \mu_{\text{B}}) \cdot h = 0.07h\cdot \text{T}\). Thus, achieving the high-efficiency regime requires a magnetic field just below $1\text{T}$, which is experimentally available.

\section{The recursive Green's function}

\begin{figure}[htbp]
\includegraphics[width=0.8\columnwidth]{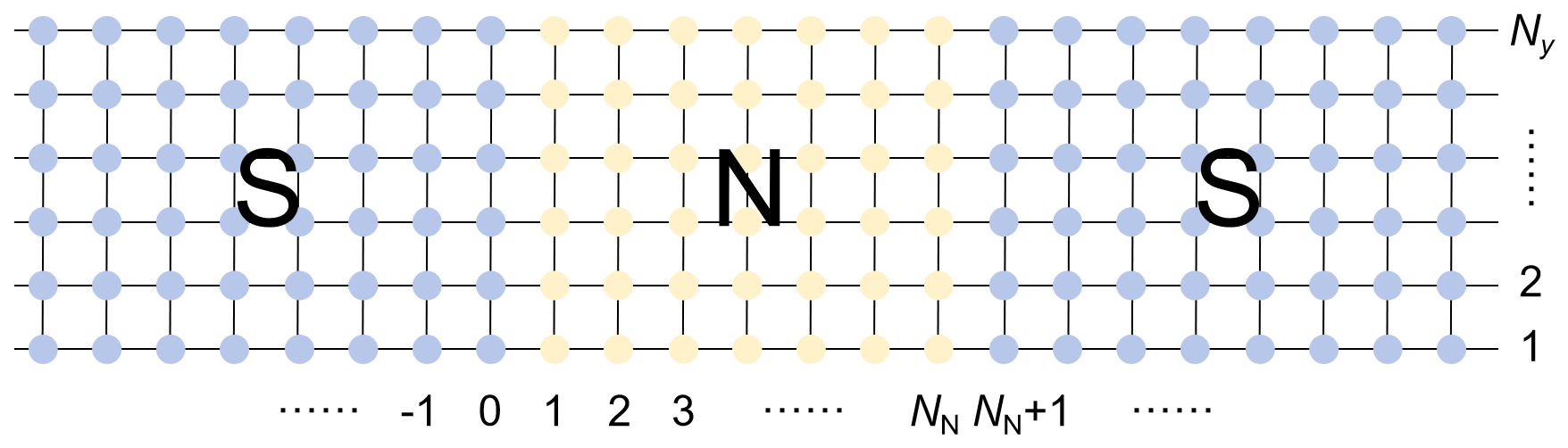}
\caption{\label{FigS1} The tight binding model for the multiband nanowire. Here $N_y$ denotes the number of orbital channels included in the system. Each orbital is treated as a chain along the $x$-direction. The coordinate origin in the $x$-direction is defined at the interface between the left lead and the normal region.}
\end{figure}

We present the details of the recursive Green's function method~\cite{SDE_2024_PRL_QFSun} employed to compute the supercurrent in the main text. The continuum system is discretized into a multichain tight-binding model, as schematically illustrated in Fig.~\ref{FigS1}. The system is partitioned into on-site terms, \( H_{0,\text{L}/\text{R}/\text{N}} \), and coupling terms, \( H_x \) and \( H_y \). The on-site Hamiltonian consists of the left superconducting lead (\( H_{0,\text{L}} \)), the central normal region (\( H_{0,\text{N}} \)), and the right superconducting lead (\( H_{0,\text{R}} \)), each comprising \( N_y \) tight-binding chains, where \( N_y \) corresponds to the number of orbital channels considered. The chains are coupled via spin-orbit interactions described by \( H_y \), while adjacent sites along the \( x \)-direction are connected through \( H_x \). The explicit forms of these terms are given below.
\begin{align}
H_{0,\text{L}} &= -\mu_s  \sigma_0 \otimes\tau_z + h_x\sigma_x\otimes\tau_z+h_y\sigma_y\otimes\tau_0 + \Delta \sigma_y \otimes \tau_y  \nonumber \\
H_{0,\text{N}} &= -\mu_n \sigma_0 \otimes\tau_z + h_x\sigma_x \otimes\tau_z + h_y\sigma_y \otimes\tau_0 \nonumber \\
H_{0,\text{R}} &= -\mu_s \sigma_0 \otimes\tau_z + h_x\sigma_x\otimes\tau_z+h_y\sigma_y\otimes\tau_0 + \Delta \left( \cos\phi \cdot\sigma_y \otimes \tau_y +\sin\phi \cdot\sigma_y \otimes \tau_x \right)  \nonumber \\
H_x &= -t_x \sigma_0 \otimes \tau_z + \text{i} \alpha_x \sigma_y \otimes \tau_z \\ \nonumber
H_y &= \alpha_y \sigma_x \otimes \tau_0
\end{align}
Here, $\sigma$ denotes the spin degree of freedom, while $\tau$ labels the particle-hole space. The three regions differ only in the superconducting term \( \Delta \): it vanishes in the central normal region (\( \Delta = 0 \)) but is finite in the left and right topological regions. A superconducting phase difference $\phi$ exists between the left and right superconductors. All other parameters, including the hopping term, magnetic field, and spin-orbit coupling, remain identical across the three regions, as they are part of the same nanowire. Consequently, the spin-orbit coupling is uniform along both the \( x \)- and \( y \)-directions.

In Fig.~\ref{FigS1}, the system extends significantly in the \(x\)-direction, while its size in the \(y\)-direction is restricted to \(N_y\) sites. For simplicity, each column is treated as a single unit, with the Hamiltonian of an individual column denoted as \(H_{00,\alpha}\), where \(\alpha=(\text{L},\text{N},\text{R})\) represents the left superconducting region, the normal region, and the right superconducting region, respectively. The coupling between neighboring columns in the \(x\)-direction is described by \(H_{01,\alpha}\) with $\alpha=(\text{L},\text{N},\text{R})$. Notably, \(H_{01,\alpha}\) is identical across these regions, as the coupling in the \(x\)-direction involves only hopping and spin-orbit coupling, which remain uniform throughout the nanowire. Then the Hamiltonians could be rewritten as
\begin{align}
H_{00, \alpha} =
\begin{bmatrix}
H_{0, \alpha} & H_y &  &  &  &  \\
H^*_y & H_{0,\alpha} & H_y &  &  &  \\
 & H^*_y & H_{0,\alpha} & H_y &  &  \\
 &  &  & \ddots &  &  \\
 &  &  & H^*_y & H_{0, \alpha} & H_y \\
 &  &  &  & H^*_y & H_{0,\alpha}
\end{bmatrix},
\qquad
H_{01,\alpha} =
\begin{bmatrix}
H_x & 0 &  &  &  &  \\
0 & H_x & 0 &  &  &  \\
 & 0 & H_x & 0 &  &  \\
 &  &  & \ddots &  &  \\
 &  &  & 0 & H_x & 0 \\
 &  &  &  & 0 & H_x
\end{bmatrix}
\end{align}

The current can be derived from the Green's function
\begin{align}
I & = -e \left\langle \frac{\mathrm{d}\hat{N}_\text{L}}{\mathrm{d}t} \right\rangle = \frac{\text{i}e}{\hbar} \left\langle \left[\sum_{i_x<0,\sigma} \hat{c}_{i_x,\sigma}^\dagger \hat{c}_{i_x,\sigma}, H\right] \right\rangle \\ \nonumber
&= \frac{\text{i}e}{\hbar} \sum_\sigma (t_0 \langle \psi^\dagger_{0,\sigma} \psi_{1,\sigma}\rangle - t_0 \langle \psi^\dagger_{1,\sigma} \psi_{0,\sigma} \rangle ) \\ \nonumber
&=\frac{e}{\hbar} \mathrm{Tr} \left\{ \Gamma_z \left[ {G}_{\text{10}}^<(t,t) \hat{T}-\hat{T}^\dagger {G}_{\text{01}}^<(t,t) \right] \right\}.
\end{align}
We adopt $\Gamma_z = \text{diag}(1,1,-1,-1)$ and transform the lesser Green's function into energy space according to ${G}_{NL}^{<}=\int\mathrm{d}\epsilon {G}^<(\epsilon)/2\pi$, where $\epsilon$ denotes the energy. In the energy representation, the lesser Green's function is given by ${G}^< (\epsilon) = -f(\epsilon) [{G}^r(\epsilon)-{G}^a(\epsilon)]$, with $f(\epsilon)$ being the Fermi distribution function. The retarded Green's functions, $G_\text{01}^r, G_\text{10}^r$,  are obtained via the Dyson equations $G_\text{01}^r = g_\text{00}^r\Sigma_\text{01}^r G_\text{11}^r,G_\text{10}^r=G_\text{11}^r\Sigma_\text{10}^r g_\text{00}^r$, with $\Sigma_\text{01}^r = \hat{T},\Sigma_\text{10}^r = \hat{T}^\dagger$. Here, $g_\text{00}^r$ is the surface Green's function of the left lead. ${G}_\text{11}^r$ is the Green's function of the leftmost layer $(i_x=1)$ of the normal region, calculated by a recursive algorithm~\cite{SDE_2024_PRL_QFSun, Sunqingfeng2016PRB}. The advanced Green's function satisfies  ${G}_\text{10}^a=({G}_\text{01}^r)^\dagger,{G}_\text{01}^a=({G}_\text{10}^r)^\dagger$.  Then we need to derive the surface Green's function. Based on the Hamiltonian derived above, the Green function satisfies a recursive equation shown below,
\begin{align}
\begin{pmatrix}G_{n+2,0}\\G_{n+1,0}\end{pmatrix}=\begin{pmatrix}H_{01}^{-1}(EI-H_{00})&H_{01}^{-1}H_{10}\\I&0\end{pmatrix}\begin{pmatrix}G_{n+1,0}\\G_{n,0}\end{pmatrix}=\hat{T}\begin{pmatrix}G_{n+1,0}\\G_{n,0}\end{pmatrix}
\end{align}
we  know that \( G_{n,0} \) in the limit \( n \to \infty \) must approach to zero. This means that the powers of the matrix on the right also approach zero. In other words, \( G_0 \) is composed of the eigenvectors of \( \hat{T} \) corresponding to eigenvalues smaller than 1, which means
\begin{align}
\binom{G_{1,0}}{G_{0,0}}=\binom{S_1}{S_2}A,
\end{align}
where \( \binom{S_1}{S_2} \) is the eigenvector matrix of \( \hat{T} \) corresponding to eigenvalues smaller than 1. From this, we quickly obtain
\begin{equation}
    G_{00}=(EI-H_{00}-H_{01}S_1S_2^{-1})^{-1}.
\end{equation}
Using the surface Green's function obtained, we employ a recursive algorithm to determine the retarded Green's function for the leftmost column of the normal region. The Green's function \( G_i \) for each column satisfies the recursive relation
\begin{align}
G_i = (EI-H_{00}-H_{10}G_{i-1}H_{01})^{-1}.
\end{align}

\section{High-Efficiency Plateau with Different Other Parameters}

\begin{figure}[htbp]
\includegraphics[width=0.55\columnwidth]{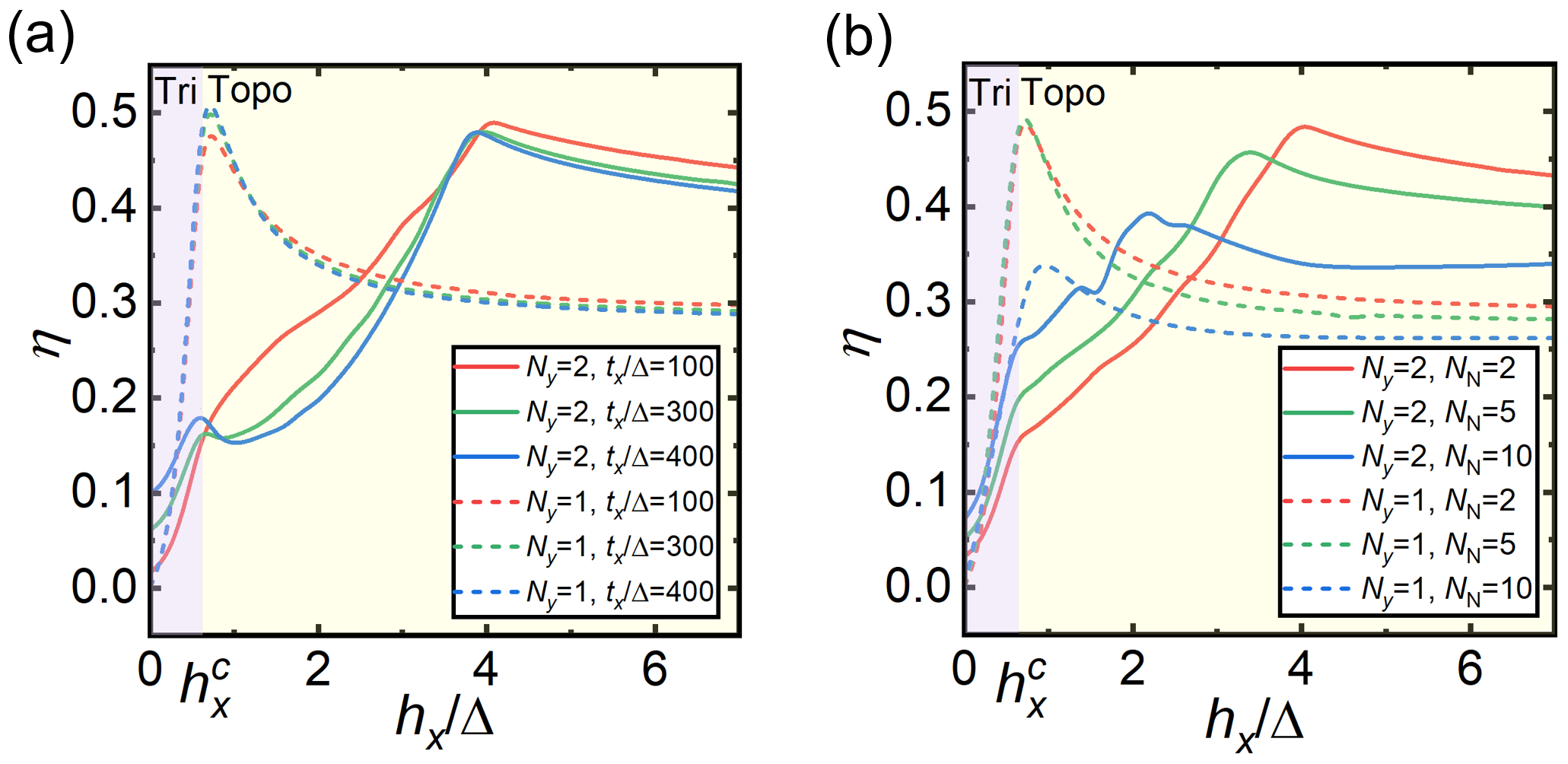}
\caption{\label{FigS2} The diode efficiency $\eta-h_x$ curves of the 2-orbit model with 3 subbands partially filled under different $t_x$'s (a) and for the diverse $N_{\text{N}}$'s (b). Parameters: $t_x=200, \,\alpha_x=30,\,V=2.5,\,h_y=0.8$.}
\end{figure}

In this section, we demonstrate that the high-$\eta$ plateau at large $h_x$ is observed across a wide range of $t_x$ and $N_{\text{N}}$ values, providing further evidence of its universality. As shown in Fig.~\ref{FigS2}(a), the efficiency curve remains almost unaffected by variations in $t_x$. This insensitivity can be attributed to the facts that the orientations of spins at low energy points (like $k_x=k_\textrm{F}\sim2\alpha_x/(t_xa_x)$ and $k_x=0$) are almost invariant under changes of $t_x$, and the bouncing mode(s) within the normal region of a relatively short Josephson junction is approximately irrelavant to the effective mass. As a result, both the MBSs and the ABSs are insensitive to altering of $t_x$ here. 
Additionally, Fig.~\ref{FigS2}(b) shows that this enhancement persists even when the junction length reaches $\sim 10a_x \sim 40 \, \text{nm}$. The observed reduction in the efficiency plateau for larger $N_{\text{N}}$ can be explained by differing decay rates: while MBSs exhibit exponential decay, ABSs decay polynomially~\cite{ABS_2018}, leading to a diminishing effect as $N_{\text{N}}$ increases.

\section{Decline of the Efficiency Plateau under Inappropriate Subband Filling}

\begin{figure}[htbp]
\includegraphics[width=1.0\columnwidth]{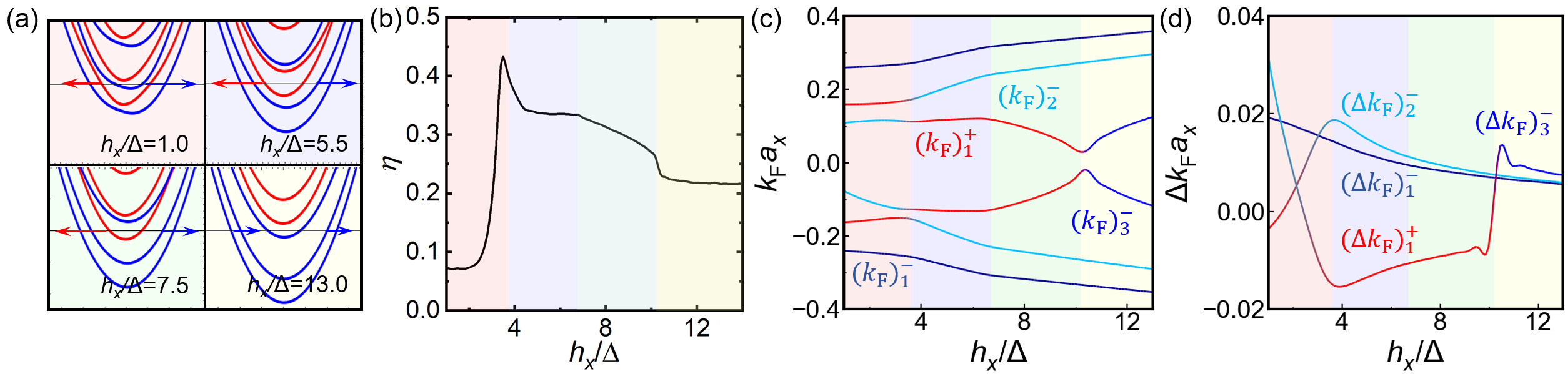}
\caption{\label{FigS3} Cases of 3-orbit model with 3 subbands occupied. (a) The dispersion relations under different $h_x$'s. (b) The diode efficiency $\eta$ versus $h_x$. (c)-(d) The curves of the Fermi wavevectors and the shifts in the Fermi points as functions of $h_x$. Parameters: $t_x=200, \,\alpha_x=10,\, V=2.5,\, \alpha_y=1.3,\, h_y=0.8$.}
\end{figure}

As shown in Fig.~\ref{FigS3}(b), when a 3-orbit ($N_y=3$) model with 3 subbands populated (more realistic and reasonable than the model of $N_y=2$ model with 3 subbands occupied) is adopted, the left part of the efficiency curve $\eta-h_x$ is almost the same as that in Fig.~\ref{Fig3}(b). This is reasonable, since the subbands with $i_y=3$ are high-energy and far away from the Fermi surface, causing weak effect to the 2-orbit part, as illustrated by the band dispersions in Fig.~\ref{FigS3}(a). The shifts of Fermi points also make a proof, where Fig.~\ref{Fig4}(a) and the left part of Fig.~\ref{FigS3}(c) ($h_x<6$) share the same feature.

When the parallel component of external magnetic field $h_x$ continues to enlarge, the high efficiency plateau will turn into a slope downwards in the middle of the blue region. This decline happens when the top blue subband and the middle red one exchanges, and arises from the change of speed of Fermi points' shifting.

As $h_x$ keeps increasing, there exists the second critical value, where a sudden drop instead of rise in $\eta$ occurs. At this time, the lowest red subband and the highest blue one pass through each other, while the Fermi points closest to $k_x=0$ suddenly encounter spin-parity flip. After that, the $\eta-h_x$ curve enters the relative final yellow region and performs as a relatively low platform, which arises from the imbalance of bound states: both ABSs and MBSs come from red subbands with the same shifting direction. This can be directly shown by the right part in Fig.~\ref{FigS3}(c) ($h_x>10$).

In conclusion, this is an example of inappropriate subband filling under a relatively large magnetic field. This provides a counter-argument to the conjecture we made in the text, that it may be better to populate $2N_y-1$ spinful subbands when $N_y$-orbit model is necessary and enough under a certain $h_x$. If a specific number of subbands are occupied but the $h_x$ is constantly increased, the high-efficiency platform will not extend infinitely.

\section{Fermi Point Shifting for Model with More Orbits}

\begin{figure}[htbp]
\includegraphics[width=0.55\columnwidth]{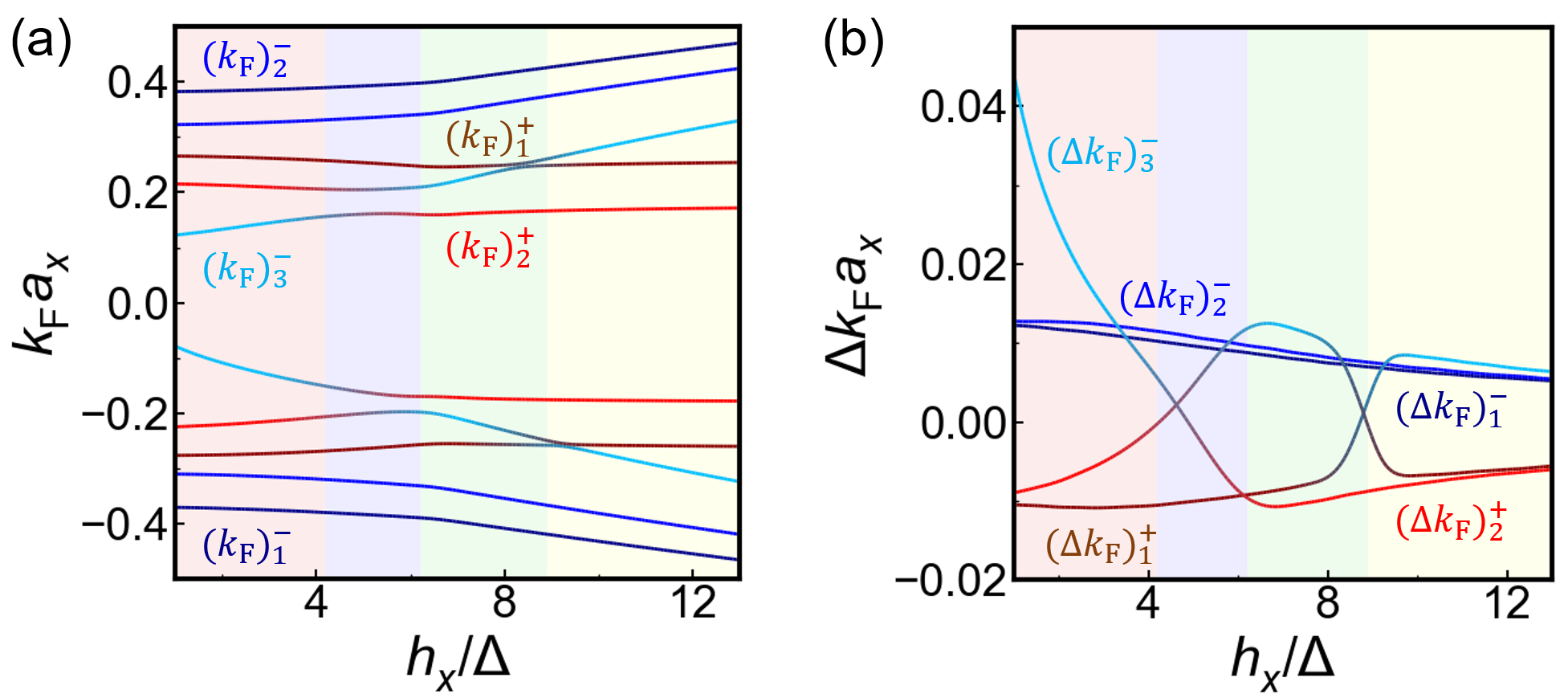}
\caption{\label{FigS4} (a) Fermi vectors with and without $h_y$ for $N_y=3$ with 5 subbands populated. (b) Shifts of Fermi Points. Parameters: $t_x=200, \,\alpha_x=10,\, V=2.5,\, \alpha_y=1.3,\, h_y=0.8$.}
\end{figure}

In this section, we detail how the diode efficiency is controlled by adjusting the magnetic field $h_x$ for the case of $3$-orbital degree of freedom and $5$ subband occupied. As shown in Fig.~\ref{FigS4}(a), five pairs of Fermi points are plotted, and their positions vary as $h_x$ increases. This process reveals two distinct spin-parity subband exchange regions, since there is always $|k_{\textrm{F}}|^-_2>|k_{\textrm{F}}|^+_1$ here because of adequate SOI, avoiding one exchange. Correspondingly, in Fig.~\ref{FigS4}(b), the Fermi point shift \( \Delta k_x a_x \) crosses zero twice, indicating band exchanges happens in $h_x\approx 5$ and $h_x\approx9$.
After the second exchange, the 4 Fermi points with smallest wavevectors shift in the same direction when $h_y$ is applied, while the other 6 points shift oppositely. Consequently, the top 2 occupied subbands contribute better due to the nonlinear dispersion, and ABSs can balance MBSs better.

\section{Band topology and exchange condition in multiband Josephson diode}

The topology of our system can be determined by either evaluating winding number, or examining possible gap closures. Since our system is quasi one-dimensional, we can adopt the algorithm invented by Fukui, Hatsugai and Suzuki ~\cite{doi:10.1143/JPSJ.74.1674}. After discretizing the Brillouin zone into {$k_1$,$k_2$,$\dots$,$k_s$} and constructing the tensor element $U_{m,n}^{(i)}=\braket{\psi_m(k_{i+1})|\psi_n(k_i)}$ for each $k_i$, where $\psi_l(k_i)$ is the $l$-th eigenvector of the Hamiltonian $H(k_i)$, then one can numerically calculate the winding number $\nu$ by
\begin{equation}
\nu=\frac{1}{\pi} \operatorname{Im}\left[\log \left(\prod_{i=1}^s \frac{\left|U^{(i)}\right|}{\sqrt{\left|U^{(i)}\right|\left|U^{(i)}\right|^*}}\right)\right].
\end{equation}
Here, $|\cdot|$ denotes the determinant of the tensor U, and periodic boundary condition should be adopted.

For the ideal 1-orbit model, the criterion for topological phase transition has already been found~\cite{2024_PRB_JingWang_Ideal1DJDE}:
\begin{equation}
    \mu_0^2+\Delta^2=(h_x^c)^2+h_y^2.
\end{equation}
For the 2-orbit case, since $|\mu_1-\mu_2|\gg\Delta$ in experiment and $h_y<\Delta$, when ignoring $\alpha_y$, we can find that at $k_x=0$, as $h_x$ exceeds $h_x^c\approx\sqrt{\Delta^2-h_y^2}$ and keeps increasing, there will be no band closure. Therefore, no topological phase transition will occur subsequently. Considering $\alpha_y$, it only affects the position of $h_x^c$ to a certain extent, but does not significantly influence the value of the starting point $h_x^{c'}\approx|\mu_1-\mu_2|/2$ of the high-efficiency platform.


\end{document}